\documentclass{iau} 
\usepackage{graphicx}

\title[SuperMALT Survey: Evolution of massive and dense clumps] 
{SuperMALT: Physical and Chemical Properties of Massive and Dense Clumps}

\author[Neupane S., Garay G. \& Contreras Y. et al. ]   
{Sudeep Neupane$^1$, Guido Garay$^1$, Yanett Contreras$^2$,
 \and SuperMALT Team$^3$}

\affiliation{$^1$Department of Astronomy, Universidad de Chile, Casilla 36D, Santiago, Chile \\ Email: {\tt sneupane@das.uchile.cl} \\[\affilskip]
$^2$Leiden Observatory, Leiden University, PO Box 9513, NL-2300 RA Leiden \\[\affilskip] $^3$ http://home.strw.leidenuniv.nl/$\sim$ycontreras/supermalt.html}

\pubyear{2017}
\volume{332}  
\jname{Astrochemistry VII- Through the Cosmos from Galaxies to Planets}
\editors{A.C. Editor, B.D. Editor \& C.E. Editor, eds.}
\begin{document}

\maketitle

\begin{abstract}
The SuperMALT survey is observing 76 MALT90 clumps at different evolutionary stages (from pre-stellar or quiescent to HII) in high excitation molecular lines and key isotopomers using the Apex 12m telescope with an angular resolution of $\sim$20'' and a velocity resolution of $\sim$0.1 km/s. The aim of this survey is to determine the physical, chemical, and kinematical properties of the gas within clumps as they evolve. Here we report some preliminary results based on observations of the  $J$=3-2 \& 4-3 lines of HNC, HCN, HCO$^+$, N$_2$H$^+$ and of the  $J$=3-2 line of the isotopologue H$^{13}$CO$^+$. We find that the morphologies and line profiles vary with the evolutionary stage of the clumps. The average line width increases from quiescent to HII clumps while line ratios show hint of chemical differences among the various evolutionary stages.

\end{abstract}

\firstsection 
\section{Introduction}

Several observations have shown that high-mass stars form in massive dense clumps which have typically masses of $\sim$ 10$^{3}$ M$_{\odot}$, densities of $\sim$10$^{5}$ cm$^{-3}$ and dust temperatures $\le$ 20 K. (eg., \cite[Faundez et al. 2004]{Faundez2004}, \cite[Garay et al. 2004]{Garay2004}). Despite of the great role that high-mass stars play as energy input in the Galaxy, their formation mechanism is not cleary understood. Determining the properties of massive dense clumps and their fragmentation process, giving rise to cores,  is crucial to understand  the formation process of high-mass stars.
Several molecular and continuum surveys have been made to investigate the characteristics and evolutionary stages of massive dense clumps (eg., ATLASGAL: \cite[Schuller et al. 2009]{Schuller2009}, Hi-GAL: \cite[Molinari et al. 2010]{Molinari2010}, GLIMPSE/MIPSGAL: \cite[Benjamin et al. 2003]{Benjamin2003}; \cite[Carey et al. 2009]{Caray2009}, MALT90: \cite[Foster et al. 2011]{Foster2011}; \cite[Jackson et al. 2013]{Jackson2013}). In particular, the Millimeter Astronomy Legacy Team 90 GHz (MALT90) Survey observed $\sim$3200 dense molecular clumps in different evolutionary states -- from pre-stellar, to proto-stellar, and  H II regions -- in 16 low excitation molecular lines (mostly $J$=1$\rightarrow$0), near 90 GHz. This survey, made using the 22m Mopra Telescope with an angular resolution $\sim$40'' and a velocity resolution of $\sim$ 0.1 km/s, has provided important information concerning the morphology and  kinematics of the massive dense clumps showing that clumps in different evolutionary stages indeed show different physical and kinematical properties (eg., \cite[Guzman et al. 2015]{guzman2015}, \cite[Rathborne et al. 2016]{jill2016}, \cite[Contreras et al. 2017]{contreras2017}). However, since for a given molecular species only a single rotational transition (mostly $J$=1$\rightarrow$0) was observed, the  MALT90 data are insufficient to perform a robust derivation of the gas properties. Only with multi-level excitation analysis a robust determination of clump temperatures, densities, and column densities can be obtained, which are important to understand the differences in physical and chemical properties of clumps at different evolutionary stages. 

We conducted the SuperMALT survey to observe a subsample of MALT90 clumps in high excitation molecular lines and key isotopomers using the Apex 12m telescope with the aim of providing an important new legacy database to characterise the evolution of high-mass star forming clumps (\cite[Contreras et al. in preparation]{Yanett2017}). In this work, we present some preliminary results from the survey. 

\section{Observations}
We observed a subsample of 76 MALT90 clumps in different evolutionary states (pre-stellar or quiescent (Q), proto-stellar (P) and HII clumps (HII)) in high excitation molecular lines and key isotopomers (see Table \ref{tab1}). The sample was selected using the following criteria: (1) reliable detection ($\ge$5$\sigma$) of key molecular transitions (i.e. N$_{2}$H$^{+}$, HCN, HNC, HCO$^{+}$(1-0)) in MALT90; (2) masses between 400 to 10$^{4}$ M$_{\odot}$; (3) distances between 3 to 5 kpc; and (4) representative of various evolutionary stages. We used the  APEX telescope  to observe the  $J$=3-2 \& 4-3 lines of HNC, HCN, HCO$^+$, N$_2$H$^+$ and the  $J$=3-2 line of the isotopologue H$^{13}$CO$^+$  with angular resolution of $\sim$23'' at 270 GHz. As receivers we use the Swedish Heterodyne Facility Instrument (SHeFI) Apex-1 (213-275 GHZ) and Apex-2 (267-378 GHz),  with a velocity resolution of $\sim$0.1 km/s.

In the first phase of the project (2013-2016) we observed the $J$=3$\rightarrow$2 transitions and in second phase, in progress, we are observing the $J$=4$\rightarrow$3 transitions. We use the GILDAS software package to reduce and analyse the data. The CLASS and GREG package are utilised to produce the spectra, subtract baselines, fit profiles and make maps. The  final spectra for analysis have a velocity resolution 0.2 km s$^{-1}$ in $12 \times12$ pixels with a pixel size of  10''x10''. The noise in brightness temperature (T*$_{rms}$) typically is $\sim$0.03 K per channel for Apex1 observations and $\sim$0.1-0.4 K per channel for Apex2 observations depending on the molecular lines. The peak spectra  are obtained by averaging the data over a central circular region of size similar to the beam-size i.e. $\sim$20''.
In the following section we report some early results on morphology, line profiles and line ratios mainly based on $J$ 3$\rightarrow$2 transition line data.
\begin{table}[t!]
\centering
\caption{Molecular lines observed in SuperMALT. The OTF maps are 2'x2' in size.}
\label{tab1}
\begin{tabular}{l c c c}\hline
{Species} & {Frequency}  & {Observing} & {Tracer}\\ \smallskip  \smallskip
{} & {(MHz)} & { Mode} & {}    \\  \hline

HCO$^{+}$(3-2)	&	267557.633	&	OTF 	&	Density	\\
HCN(3-2)	&	265886.431	&	OTF 	&	Density	\\
HNC(3-2)	&	271981.111	&	OTF 	&	Density, Cold Chemistry	\\
N$_{2}$H$^{+}$(3-2)	&	279511.760	&	OTF 	&	Density, Chemically robust	\\
HCO$^{+}$(4-3)	&	356734.242	&	OTF 	&	Density	\\
HCN(4-3)	&	354505.473	&	OTF 	&	Density	\\
HNC(4-3)   &    362630.303            &  OTF      &   Density \\
N$_{2}$H$^{+}$(4-3)   &     372672.480            &  OTF      &   Density \\
H$^{13}$CO$^{+}$(3-2)	&	260255.342	&	Point	&	Column Density	\\
N$_2$D$^+$(3-2)	&  231321.830 & Point & Deuterium Chemistry \\ \hline
\end{tabular}
\end{table}

\begin{figure}[ht!]
\begin{center}
\includegraphics[width=0.825\linewidth]{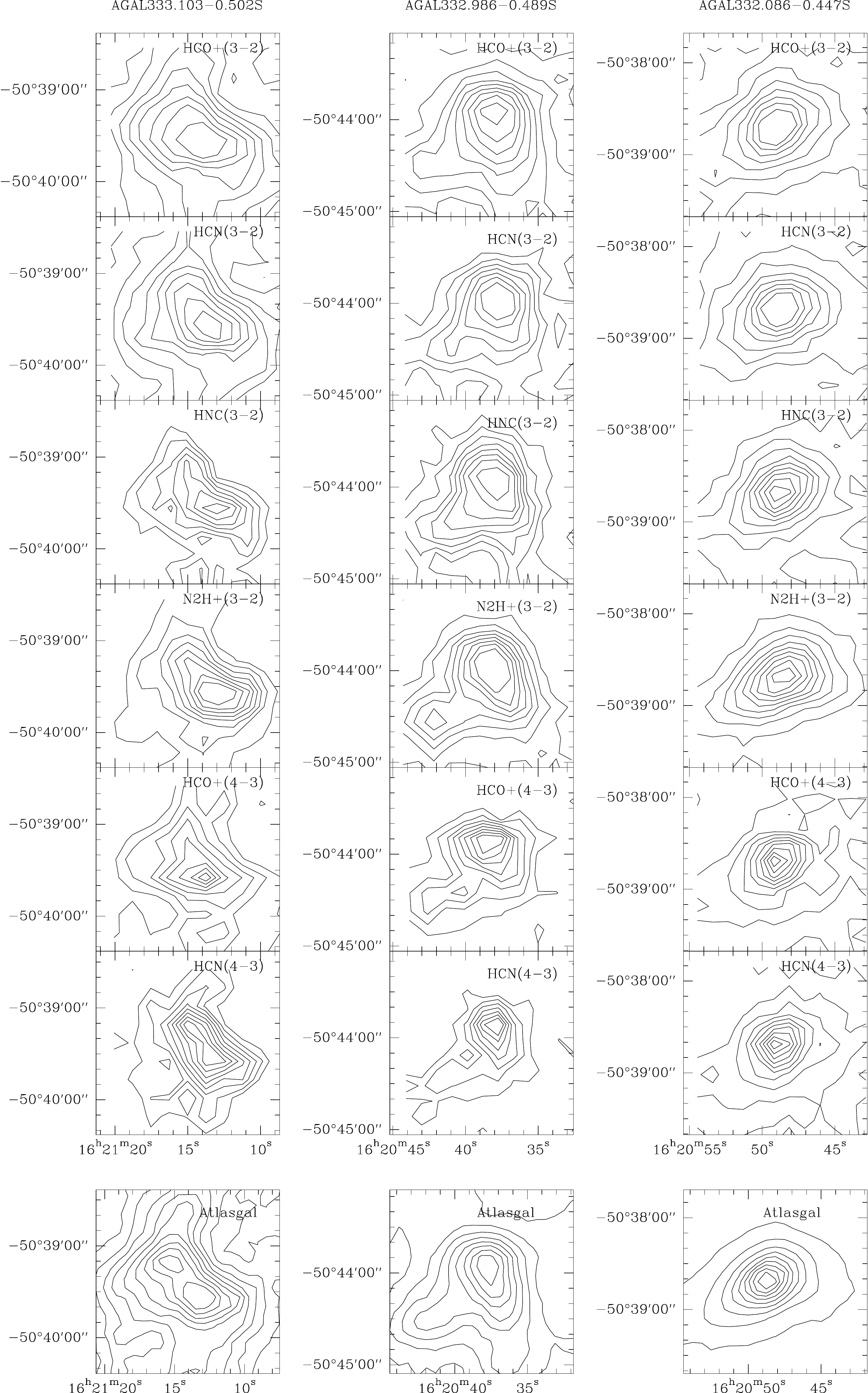} 
 \caption{Contour maps of the emissions of 6 different molecular lines and Atlasgal (870$\mu$m) continuum map from three types of clumps (AGAL333.103-0.502S: pre-stellar (left),  AGAL332.986-0.489S: proto-stellar (middle), and AGAL332.086-0.447S: HII clump (right)). Contour levels have steps of 10\% and the peak contour is at 90\% level of the peak emission. The molecular species and transition lines are shown inside each box.}
   \label{fig1}
\end{center}
\end{figure}

\section{Results}
\subsection{Morphology}
We investigated the morphology of the clumps in the different lines by making maps of the velocity integrated emission using the same velocity interval for all lines. We find that the morphologies of the emission in the $J$=3-2 lines of the different molecules are similar, although the HCO$^{+}$ (3-2) emission  is usually more extended. This is illustrated in Figure \ref{fig1} which shows maps of the emission from three clumps in different evolutionary stages in the $J$=3$\rightarrow$2 lines of HCO$^{+}$, HCN, HNC and N$_{2}$H$^{+}$ and $J$=4$\rightarrow$3 lines of HCO$^{+}$ and HCN. Also shown, for comparison, is a map of the dust continuum emission at  870 micron taken from ATLASGAL. At first order, the morphology of the molecular emissions is similar to that of the cold dust emission. The emission of $J$=4$\rightarrow$3 lines of HCO$^{+}$ and HCN trace high density regions within the clump. 

\begin{figure}[ht!]
            \centering
            \includegraphics[width=0.50\linewidth]{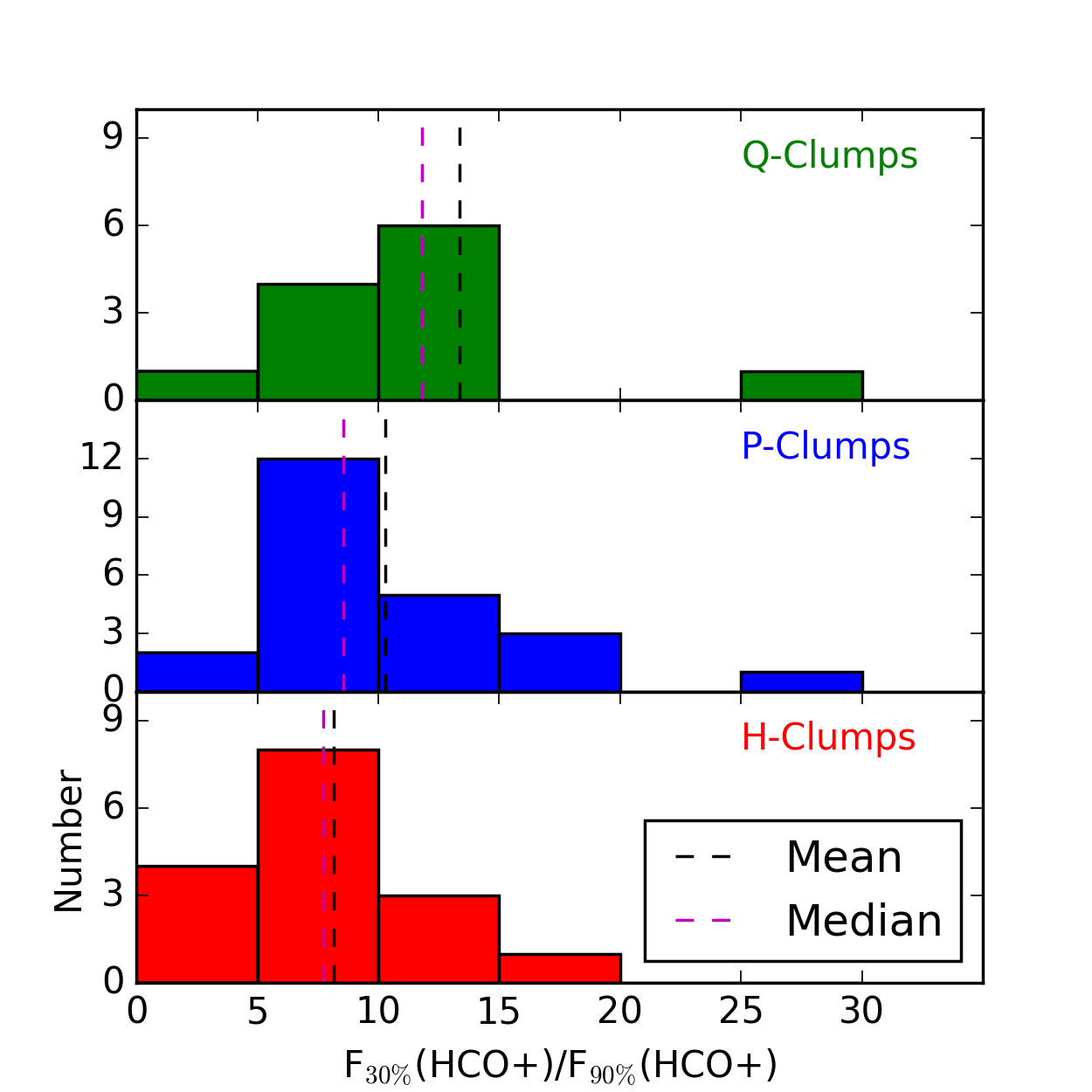}
\caption{HCO$^{+}$(3-2) integrated intensity ratios within contour region of 30\% and 90\% of the peak towards different type of clumps. The mean and median for each distribution is indicated by dotted color lines. }
\label{fig2}
\end{figure}

Using the HCO$^{+}$(3$\rightarrow$2) data, we computed flux densities over two areas, enclosing, respectively, the 30\% (F$_{30\%}$) and 90\% (F$_{90\%}$) levels of the peak emission. We find that the average value of the flux density ratio (F$_{30\%}$)/(F$_{90\%}$) is higher for quiescent clumps  than for proto-stellar and HII clumps [see Fig. \ref{fig2}] suggesting that the prestellar clumps have shallower density profile than more evolved clumps. 

\begin{figure}[ht!]
\centering
\includegraphics[width=0.83\linewidth]{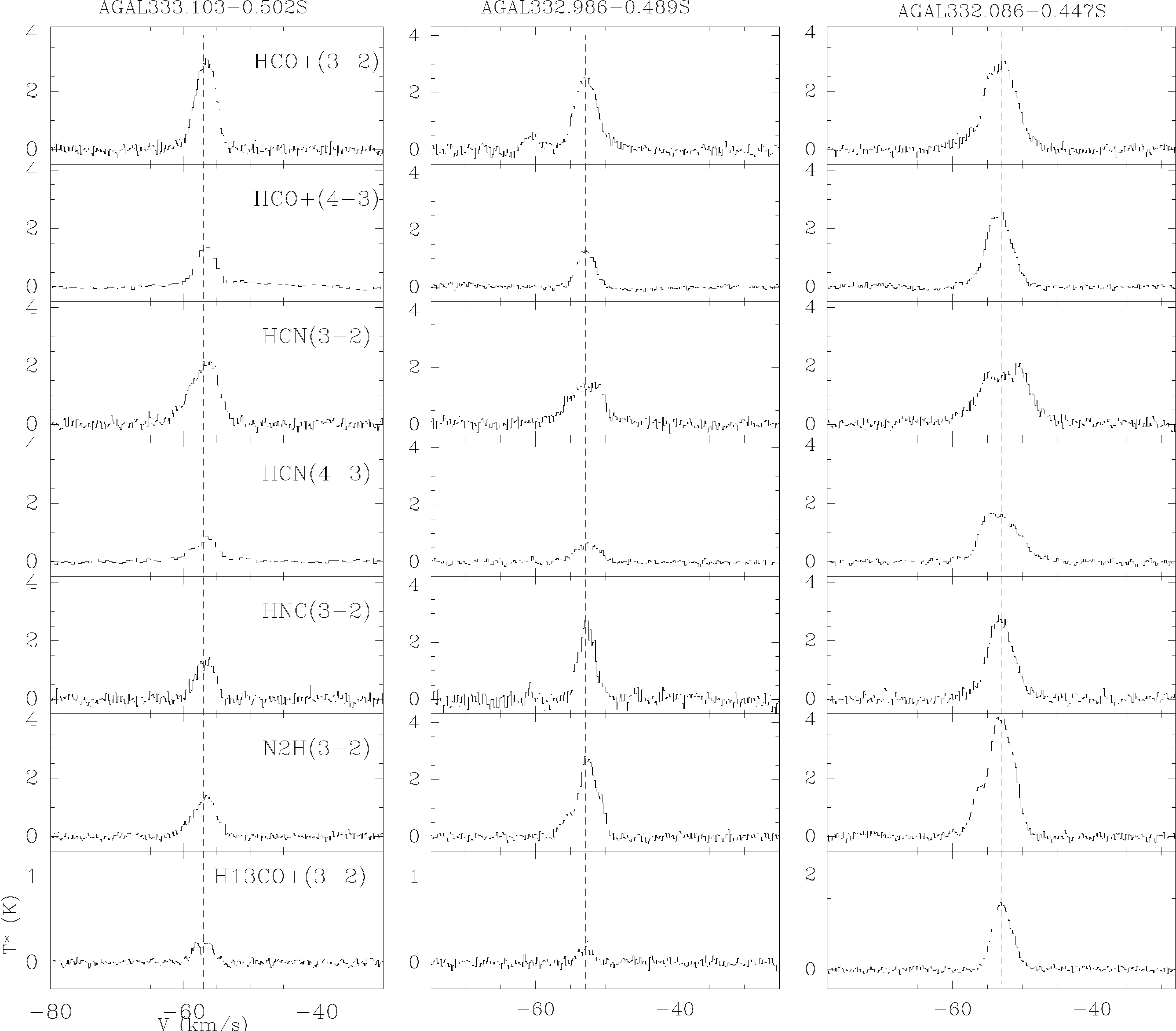}
\caption{Line profiles of 6 different molecular lines towards three different types of clumps; pre-stellar: AGAL333.103-0.502S (left), proto-stellar: AGAL332.986-0.489S (middle), HII clump: AGAL332.086-0.447S (right). The molecular species, for each row, are shown inside each box on the left.}
\label{fig3}
\end{figure}

\subsection{Line Profiles} 
The line profiles of the emission from the clumps show a variety of shapes:  gaussian, asymmetric, wings etc. The line profiles of Q-type clumps are typically nearly Gaussian, while the more evolved clumps (P and H type) show features such as self absorption and line wings [see Fig. \ref{fig3}] in optically thick lines HCO$^{+}$(3-2) and HCN(3-2). 
\begin{figure}[ht!]
            \centering
		    \includegraphics[width=0.55\linewidth, height=0.5\linewidth]{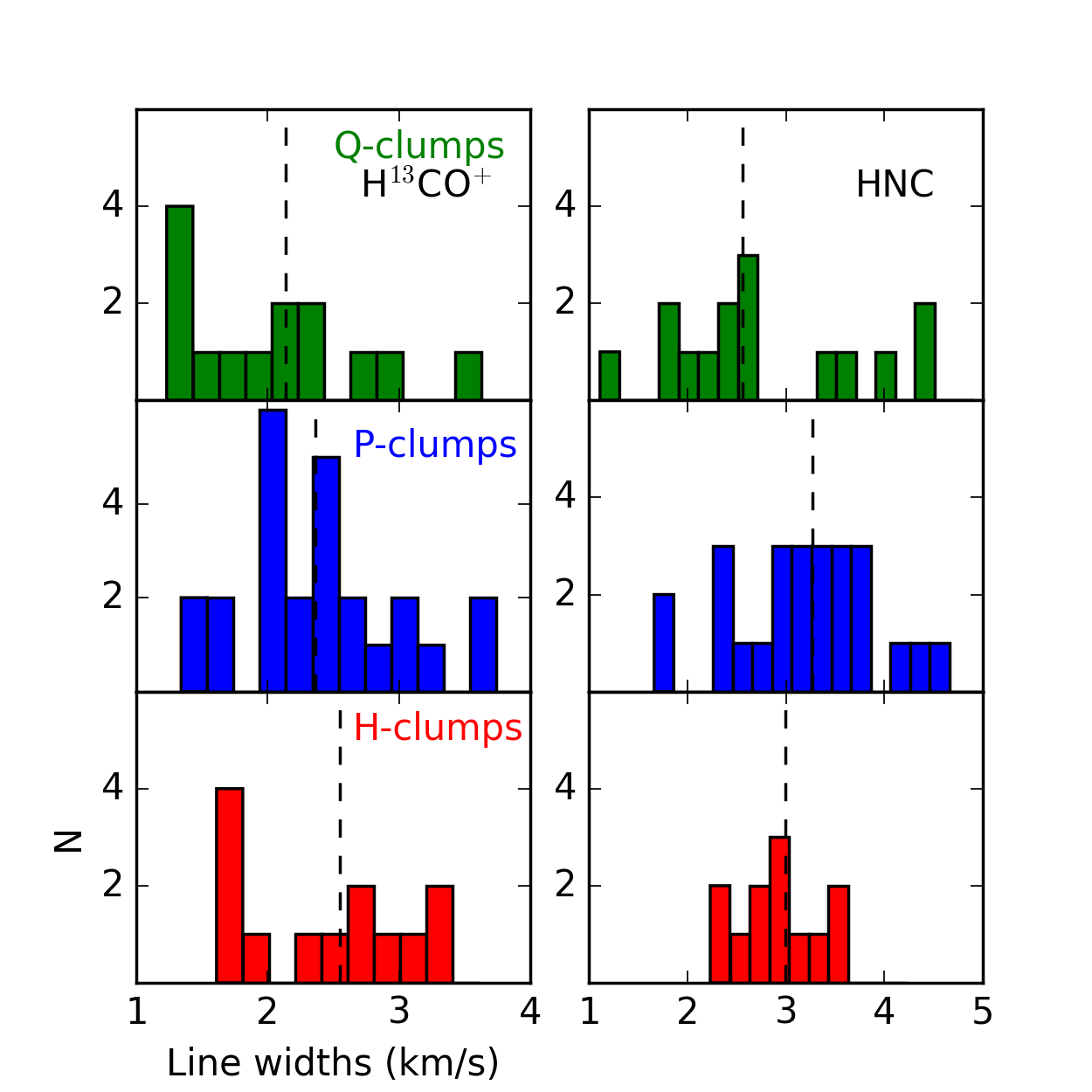}
\caption{Line widths, of Q, P and HII clumps, determined from Gaussian fit to the peak spectra of optically thin lines H$^{13}$CO$^{+}$(3-2) and HNC(3-2). The mean value for each distribution is indicated by dotted black line. }
\label{fig4}
\end{figure}

\begin{figure}[ht!]
            \centering
			\includegraphics[width=0.5\linewidth]{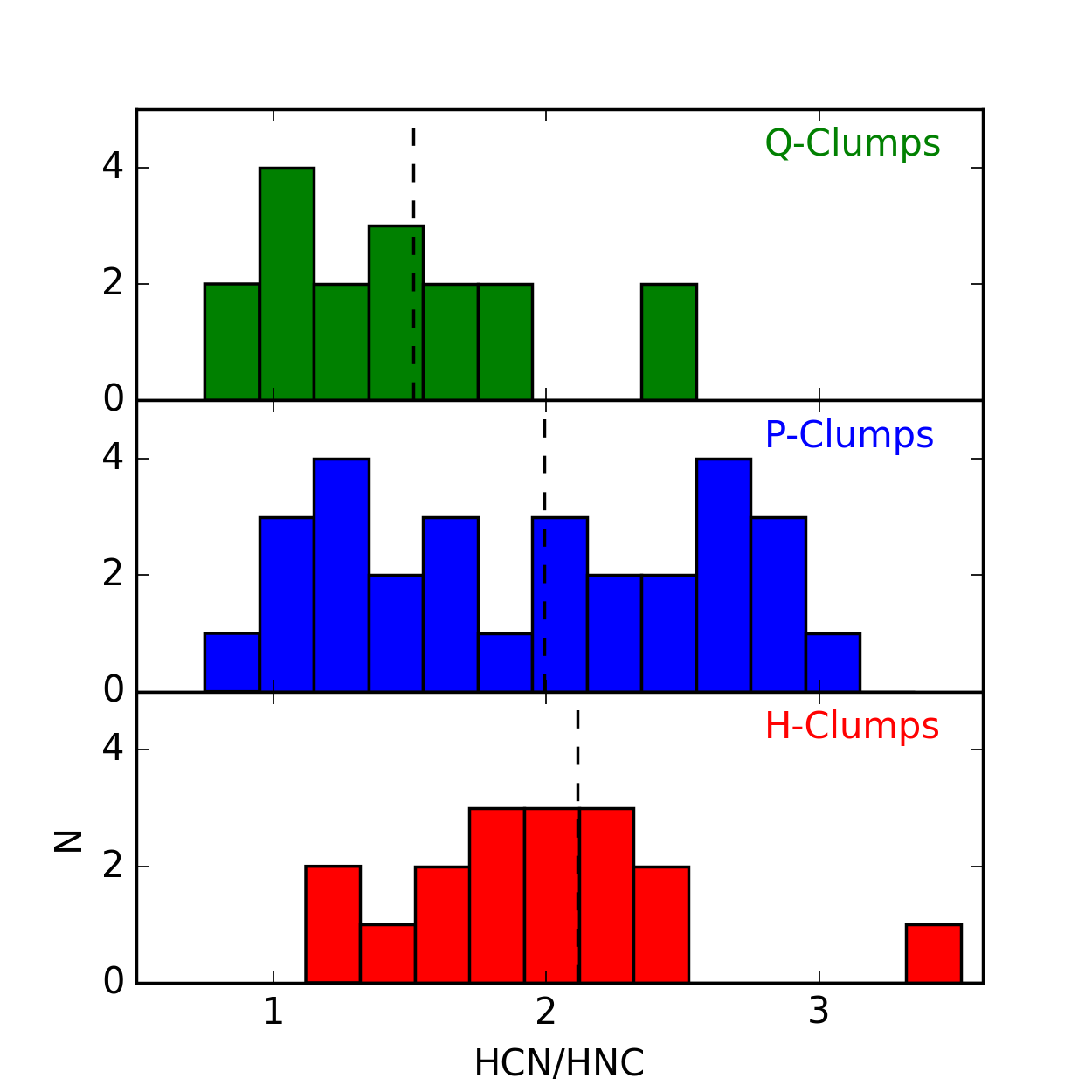}
\caption{HCN(3-2) to HNC(3-2) integrated intensity ratios from the peak spectra towards Q, P and HII clumps. The average for each distribution is indicated by dotted black line. }
\label{fig5}
\end{figure}

{\it Line width:} The width of optically thin lines  provide information about the non-thermal or turbulent velocities in the clouds. The profiles of the H$^{13}$CO$^{+}$(3-2) and HNC(3-2) emission from most clumps are Gaussian, hence we fitted gaussian profiles to obtain FWHM  widths. We used S/N $\ge$ 3$\sigma$ as detection limit for the line widths analysis of different type of clumps. We find that the average line widths increases with the evolutionary stage of the clumps [see Fig.\ref{fig4}].

{\it Line ratios:} The molecular line intensities and ratios reveal a wide range of chemical properties of clumps. Without going into complex chemistry of the molecules, we look for correlation of line intensity ratios of different species for clumps in all evolutionary stage.  The preliminary analysis shows that the HCN(3-2) to HNC(3-2) intensity ratio increases with evolutionary stage (Quiescent to HII clumps) suggesting a temperature dependency of the ratio [see Fig.\ref{fig5}]. 

\begin{figure}[ht!]
\centering
\includegraphics[width=0.5\linewidth]{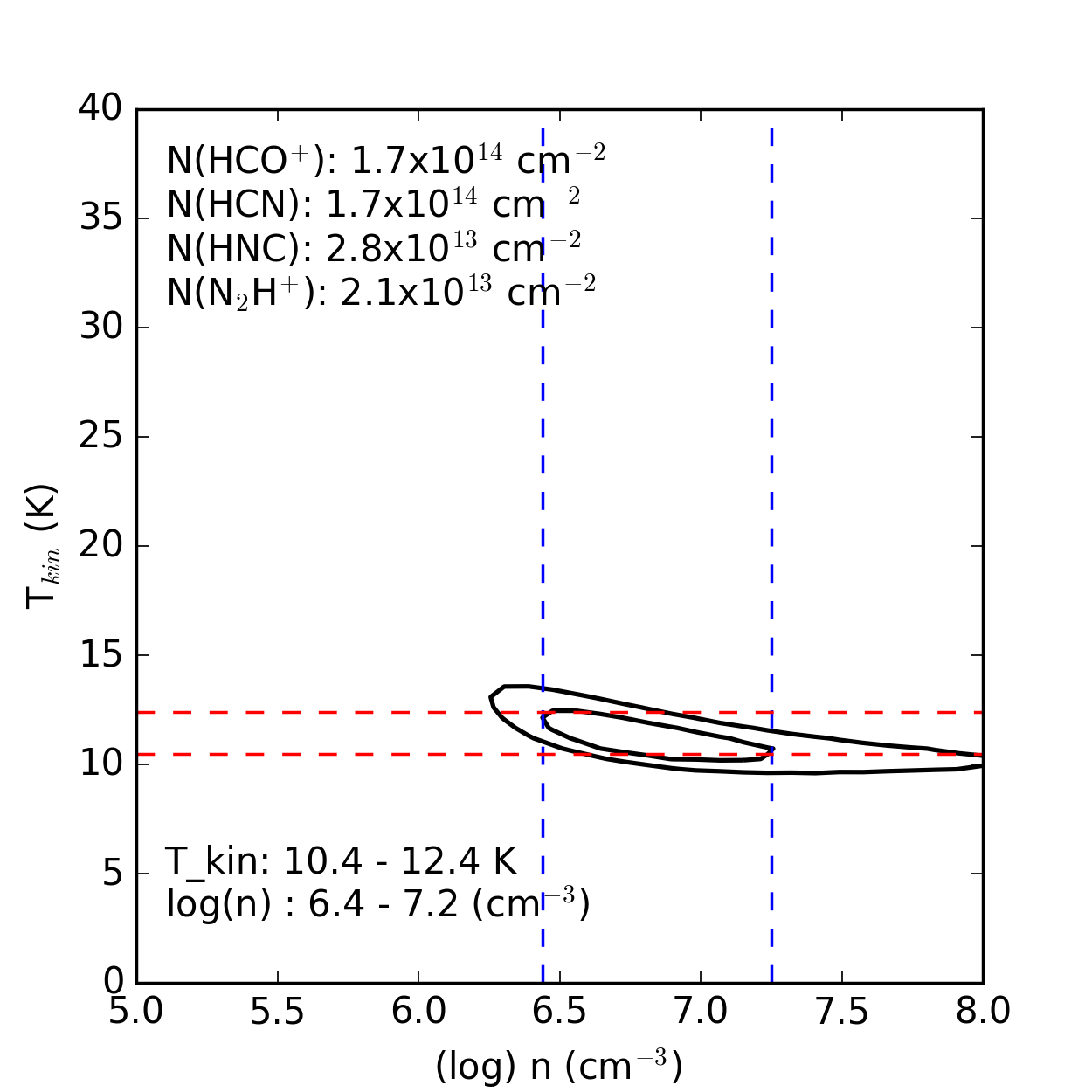}
\caption{Chi-square minimization of observed and RADEX model intensities using four main molecular lines (HCO$^{}$, HCN, HNC and N$_{2}$H$^{+}$) molecules for clump AGAL333.103-0.502S. Peak contour level shows the constraints in density and kinetic temperature.}
\label{fig6}
\end{figure}

{\it Physical properties:} The multi $J$ transition lines of molecular species can be used to constrain the physical parameters of the clumps. We use data from three rotational transitions ($J$ = 4$\rightarrow$3, 3$\rightarrow$2 from SuperMALT and $J$ = 1$\rightarrow$0 from MALT90 ) in four molecular species and perform radiative transfer modelling,  using RADEX (\cite[Schoier et al. 2005]{Schoier2005}),  to constrain the temperature and density of the clumps. This is ilustrated in Fig.\ref{fig6}, which shows a relative chi-square map made using the observed data for clump AGAL333.103-0.502S and model integrated intensities. For this clump, the kinetic temperature is constrained to be in the range 10.4 to 12.4 K and the density in the range $2.5\times10^6$ and $1.6\times10^7$ cm$^{-3}$. 

\section{Summary and Conclusions}
We report preliminary results from the SuperMALT survey. We find that the morphology of the emission observed in the $J$=3-2 lines of HNC, HCN, HCO$^+$, N$_2$H$^+$ are similar, although there are some differences. The HCO$^{+}$ emission is more extended than that seen in the N$_{2}$H$^{+}$ and HNC. The morphology of the molecular emissions are also similar to that of the dust emission at 870 micron (ATLASGAL). The line profiles appear different from quiescent clumps to HII clumps. Most quiescent clumps show nearly Gaussian line profiles while evolved clumps show features such as line wings and self absorption. Clumps in the late stages (HII)  exhibit  brighter lines than those of quiescent clumps and an increase in line widths or turbulence suggesting active star formation within the former clumps. \\

{\textit Acknowledgements:} S.N. and G.G. greatfully acknowledge support from CONICYT project PFB-06.

\end{document}